\documentclass{ws-procs9x6}

\newcommand*{\fs}[1]{{#1\!\!\!/}}
\newcommand*{\ia}{{\text{int}}}
\newcommand*{\bare}{{\text{bare}}}
\newcommand*{\ie}{\textit{i.e.}}
\begin{document}

\title{Gauge-Invariant Approach to Meson Photoproduction
Including the Final-State Interaction}

\author{H. Haberzettl}
\address{Center for Nuclear Studies, Department of Physics,\\
The George Washington University,
Washington, DC 20052, USA \\
E-mail: helmut@gwu.edu}

\author{K. Nakayama}
\address{Department of Physics and Astronomy,\\
University of Georgia,
Athens, GA 30602, USA\\
E-mail: nakayama@uga.edu}

\author{S. Krewald}
\address{Institut f\"ur Kernphysik (Theorie)\\
Forschungszentrum J\"ulich,
52425 J\"ulich, Germany\\
E-mail: s.krewald@fz-juelich.de}

\maketitle

\abstracts{A gauge-invariant formalism is presented for the practical treatment
of photo- and electroproduction of pseudoscalar mesons off nucleons that allows
an explicit incorporation of hadronic final-state interactions. The
semi-phenomenological approach is based on a field theory developed by one of
the authors. It generalizes an earlier approach by allowing for systematic
improvement of approximations in a controlled manner. The practical feasibility
is illustrated by applying the lowest-order result to the photoproduction of
both neutral and charged pions.}

%%%%%%%%%%%%%%%%%%%%%%%%%%%%%%%%%%%%%%%%%%%%%%
\section{Introduction}
The photo- and electroproduction of mesons off nucleons is one of the primary
tools to learn about the excited states of the nucleon. From a theoretical
point of view, therefore, what is needed are reaction theories that allow one
to disentangle the resonance information from the background in a reliable
manner. There exist a number of different approaches that describe the dynamics
of the photon-induced production of mesons, but the limited space of the
present contribution does not permit a detailed comparison of their respective
merits.

\begin{figure}[t!]
\includegraphics[width=\textwidth,clip=]{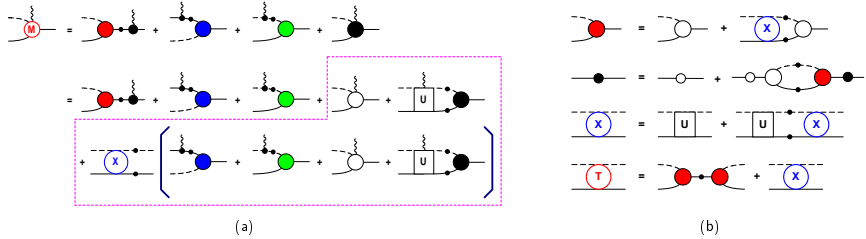}
  \caption{\label{diagrams1} Diagrammatic summary of the field-theory formalism
  of Ref.~\protect\refcite{HH1}. Time proceeds from right to left. ~(a) Meson
  production current $M^\mu$. The first line corresponds to
  Eq.~(\protect\ref{eq:Mmu_suti}) summing up, in that order, the $s$-, $u$-,
  and $t$-channel diagrams and the interaction current $M^\mu_{\rm int}$, whose
  dynamical content is exhibited by the diagrams enclosed in the dashed box of
  the last two lines. This also includes, in the bottom line, the final-state
  interaction mediated by the nonpolar $\pi N$ amplitude $X$ that satisfies the
  integral equation shown in (b). The diagram element labeled $U$ subsumes all
  exchange currents $U^\mu$ contributing to the process (see
  Fig.~\protect\ref{fig:ucurr}). The diagram with open circle depicts the bare
  current $m^\mu_{\rm bare}$ (\ie, the Kroll--Ruderman term). ~(b) Pion-nucleon
  scattering with dressed hadrons. The full $\pi N$-amplitude is denoted by
  $T$, with $X$ subsuming all of its nonpolar (\ie, non-$s$-channel)
  contributions. The latter satisfies the integral equation $X=U+UG_0X$
  depicted in the third line here, where the driving term $U$ sums up all
  nonpolar irreducible contributions to $\pi N$-scattering, \ie, all
  irreducible  contributions which do not contain an $s$-channel pole (see
  Ref.~\protect\refcite{HH1} for full details). Diagram elements with open,
  unlabeled circles describe bare quantities.}
\end{figure}

The basic feature of all approaches, however, is that topologically there are
four distinct contributions to the amplitude $M^\mu$, as shown in
Fig.~\ref{diagrams1}. There are three contributions, referred to as the $s$-,
$u$-, and $t$-channel contributions --- $M^\mu_s$, $M^\mu_u$, and $M^\mu_t$,
respectively --- according to the respective Mandelstam variables of the
intermediate hadron, where the photon attaches itself to an external leg of the
basic underlying $\pi NN$ vertex, and there is a fourth contribution, the
interaction current $M^\mu_\ia$, where the photon interacts \emph{within} the
vertex. This breakdown,
\begin{equation}
M^\mu = M^\mu_s+M^\mu_u+M^\mu_t+M^\mu_\ia~,
 \label{eq:Mmu_suti}
\end{equation}
thus is generic and reflected in all approaches. There is considerable
difference, however, in the way the four basic contributions are implemented
dynamically.

%%%%%%%%%%%%%%%%%%%%%%%%%%%%%%%%%%%%%%%%%%%%%%
%%%%%%%%%%%%%%%%%%%%%%%%%%%%%%%%%%%%%%%%%%%%%%

The present work is based on the field-theoretical approach given by
Haberzettl.\cite{HH1} The complete formalism takes into account all possible
hadronic mechanisms, including the final-state interaction (FSI), and it is
gauge-invariant as a matter of course. However, in view of its complexity and
high nonlinearity, its practical implementations require that some reaction
mechanisms need to be truncated and/or replaced by phenomenological
approximations. In doing so, and depending on the level of approximation, the
resulting reaction dynamics then become very similar to that of dynamical
meson-exchange models of hadronic interactions.

Gauge invariance is one of the fundamental symmetries that must be maintained
in any approach involving the electromagnetic interaction with hadronic matter.
Our objective here is to preserve full gauge invariance through all levels of
approximations. As a matter of course, this must also include the case where
the underlying $\pi NN$ interaction is modeled by \emph{ad hoc}
phenomenological form factors and, in particular, gauge invariance must remain
true also in the presence of explicit final-state interactions. The latter
problem has been treated already in Ref.~\refcite{HH3} as a two-step procedure
where the gauge-invariant treatment of explicit FSIs was added on to an already
gauge-invariant tree-level amplitude that had been constructed according to the
prescriptions given in Refs.~\refcite{HH2,DW}. The present approach instead
starts from the full amplitude and derives a single condition for the
mechanisms to be approximated that follows directly from the generalized
Ward--Takahashi identities given in the next section.\cite{HH1,kazes} At the
lowest order, we reproduce the earlier results.\cite{HH3} In addition, we shall
provide a general scheme that allows one, in principle, to systematically
improve on the initial approximation in a controlled manner.

In the following, for definiteness, we will explicitly consider the production
of pions off the nucleon, \ie, $\gamma+N\to \pi+N$, but the formalism will of
course apply equally well to the photo- or electroproduction of any
pseudoscalar meson. Moreover, at intermediate stages of the reaction, we will
ignore other mesons or baryonic states since they are irrelevant for the
problem at hand, \ie, how to preserve gauge invariance in the presence of
hadronic final-state interaction.

%%%%%%%%%%%%%%%%%%%%%%%%%%%%%%%%%%%%%%%%%%%%%%%%%%%%%%
\section{Gauge Invariance}

The production current $M^\mu$ is gauge invariant if its four-divergence
satisfies the generalized Ward--Takahashi identity\cite{HH1,kazes}
\begin{align}
k_\mu M^\mu & =  - [F_s\tau]S_{p+k}Q_iS^{-1}_p + S^{-1}_{p'}Q_f
S_{p'-k}[F_u\tau]
\nonumber \\
&\quad\mbox{}
 +  \Delta^{-1}_{p-p'+k}Q_\pi\Delta_{p-p'}[F_t\tau]~,
 \label{gi1}
\end{align}
where $p$ and $k$ are the four-momenta of the incoming nucleon and photon,
respectively, and $p'$ and $q$ are the four-momenta of the outgoing nucleon and
pion, respectively, related by momentum conservation $p'+q=p+k$. $S$ and
$\Delta$ are the propagators of the nucleons and pions, respectively, with
their subscripts denoting the available four-momentum for the corresponding
hadron; $Q_i$, $Q_f$, and $Q_\pi$ are the initial and final nucleon and the
pion charge operators, respectively. The index $x$ at $[F_x\tau]$ labels the
appropriate kinematic situation for $\pi NN$ vertex in the $s$-, $u$-, or
$t$-channel diagrams of Fig.~\ref{diagrams1} (and $\tau$ explicitly denotes the
isospin-operator dependence in a schematic manner for later convenience). This
is an \emph{off-shell} condition. In view of the inverse propagators appearing
in each term here, if all external hadronic legs are on-shell, this reduces to
\begin{equation}
k_\mu M^\mu =0
 \qquad \text{(hadrons on-shell)}~,
 \label{cc1}
\end{equation}
which describes current conservation.

Physically relevant, of course, is only current conservation. However, in a
\emph{microscopic} dynamical theory it is not sufficient to simply conserve the
overall current. In addition, one must have \emph{consistency} at all level of
the hierarchy of mechanisms that provide the details of the reaction dynamics.
It is found\cite{HH1,kazes} that it is \emph{not} possible to achieve current
conservation consistently unless the current satisfies the off-shell condition
(\ref{gi1}).

The electromagnetic currents for the nucleons and the pions, $\Gamma^\mu_N$ and
$\Gamma^\mu_\pi$, respectively, satisfy the Ward--Takahashi identities
\begin{subequations}\label{eq:WTpiN}
\begin{align}
k_\mu \Gamma_N^\mu(p',p) &=S^{-1}_{p'}Q_N-Q_NS^{-1}_p~,
\\
k_\mu \Gamma_\pi^\mu(q',q) &=\Delta^{-1}_{q'}Q_\pi-Q_\pi\Delta^{-1}_q~,
\label{WTN}
\end{align}
\end{subequations}
where the four-momentum relations $p'=p+k$ and $q'=q+k$ hold. It is therefore
possible to replace the generalized WT identity (\ref{gi1}) by the
\emph{equivalent} gauge-invariance condition
\begin{align}
k_\mu M^\mu_\ia & = - [F_s\tau] Q_i +Q_f [F_u\tau] +Q_\pi [F_t \tau]
\equiv - F_s e_i +F_u e_f +F_t
e_\pi~,
 \label{eq:GCMint}
\end{align}
where the operators
\begin{equation}
e_i = \tau Q_i~,
 \quad
e_f = Q_f \tau ~,
 \quad\text{and}\quad
e_\pi = Q_\pi \tau
\end{equation}
describe the respective hadronic charges in an appropriate isospin basis
(component indices and summations are suppressed here). Charge conservation for
the production process simply reads $e_i=e_f+e_\pi$. In the following, we use
the condition (\ref{eq:GCMint}), instead of (\ref{gi1}), together with
(\ref{eq:WTpiN}).

%%%%%%%%%%%%%%%%%%%%%%%%%%%%%%%%%%%%%%%%%%%%%%%%%%%%%%
\section{Formalism}

\begin{figure}[t!]
  \parbox{.6\textwidth}{%
  \includegraphics[width=.6\textwidth,clip=]{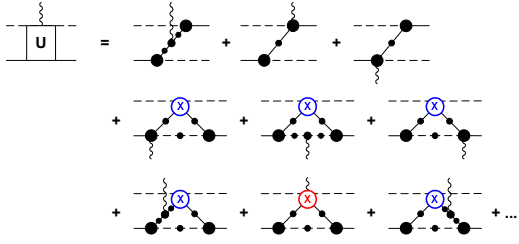}}
  \hfill
  \parbox{.33\textwidth}{%
  \caption{\label{fig:ucurr}%
  Exchange-current contributions subsumed in $U^\mu$. The three contributions
  in the top row are referred to as $E^\mu$ in the
  Eq.~(\protect\ref{eq:Eadd}).}}
\end{figure}

For the present purpose, it is sufficient to summarize the field-theory
formalism of Ref.~\refcite{HH1} in terms of the diagrams of
Figs.~\ref{diagrams1} and \ref{fig:ucurr}. It is seen here that the interaction
current $M^\mu_\ia$, where the photon couples inside the vertex, explicitly
contains the hadronic FSI. The structure of the interaction current thus is
rather complex and we read off the diagrams enclosed by the dashed box of
Fig.~\ref{diagrams1}(a) that
\begin{align}
M^\mu_\ia &= m^\mu_\bare +U^\mu
G_0[F\tau]
 \nonumber\\
 &\quad\mbox{}
+XG_0\Big(M^\mu_u+M^\mu_t+m^\mu_\bare +U^\mu G_0[F\tau]\Big)~,
\end{align}
where $m^\mu_\bare$ is the bare Kroll--Ruderman contact current, $U^\mu$
subsumes all possible exchange currents (see Fig.~\ref{fig:ucurr}), $G_0$
describes the intermediate $\pi N$ two-particle propagation, and the FSI is
mediated by the nonpolar part $X$ of the $\pi N$ $T$ matrix. The notation
$[F\tau]$ is used for the dressed initial $N\to\pi N$ vertex. The equation for
this dressed vertex is given in Fig.~\ref{diagrams1}.

The preceding expression is still exact. In practical applications, however,
one will not be able to calculate all mechanisms that contribute to the full
reaction dynamics and one must make some approximations. This is particularly
true for the complex mechanisms that enter $U^\mu$, as depicted in
Fig.~\ref{fig:ucurr}. Approximations should preserve the gauge invariance of
the amplitude. To see how this can be done, let us define
\begin{equation}
M^\mu_a = m^\mu_\bare +U^\mu G_0 [F\tau]~,
 \label{eq:Madef}
\end{equation}
thus write
\begin{equation}
M^\mu_\ia = M^\mu_a +XG_0\Big(M^\mu_u+M^\mu_t+M^\mu_a\Big)~,
 \label{eq:Mint_Ma}
\end{equation}
and recast the gauge-invariance condition (\ref{eq:GCMint}) as a condition for
$M^\mu_a$. Denoting the \emph {non-transverse} parts of the $u$- and
$t$-channel currents $M^\mu_u$ and $M^\mu_t$ by $m^\mu_u$ and $m^\mu_t$,
respectively, \ie,
\begin{equation}
 k_\mu(M^\mu_u-m^\mu_u)=0 \quad\text{and}\quad k_\mu(M^\mu_t-m^\mu_t)=0
 \label{eq:Mmdiff}
\end{equation}
holds true as an \emph{off-shell} property, this immediately leads to
\begin{align}
k_\mu M^\mu_a &= (1-UG_0)\left[-F_s e_i+F_u e_f +F_t e_\pi\right]
 -k_\mu UG_0 (m^\mu_u +m^\mu_t)
 \label{eq:GIMa2}
\end{align}
as the necessary condition that $M_a^\mu$ must satisfy so that $M^\mu_\ia$
yields the gauge-invariance condition (\ref{eq:GCMint}). Note that no
approximation has been made up to here.

%%%%%%%%%%%%%%%%%%%%%%%%%%%%%%%%%%%%%%%%%%%%%%
%%%%%%%%%%%%%%%%%%%%%%%%%%%%%%%%%%%%%%%%%%%%%%

%%%%%%%%%%%%%%%%%%%%%%%%%%%%%%%%%%%%%%%%%%%%%%%%%%%%%%
\subsection{Approximating $M^\mu_a$}

The structure of the preceding condition suggests the following approximation
strategy. The condition evidently is satisfied if we now approximate $M^\mu_a$
by
\begin{equation}
M_a^\mu = (1-UG_0) M^\mu_c - UG_0 (m^\mu_u +m^\mu_t) +T^\mu~,
 \label{eq:Ma_approx}
\end{equation}
where $M^\mu_c$ can be \emph{any} contact current satisfying
\begin{equation}
k_\mu M^\mu_c = -F_s e_i+F_u e_f +F_t e_\pi
 \label{eq:McGI}
\end{equation}
and $T^\mu$ is an undetermined transverse \emph{contact} current that is
unconstrained by the four-divergence (\ref{eq:GIMa2}). With the choice
(\ref{eq:Ma_approx}), the corresponding approximate $M^\mu_\ia$ is then easily
found from (\ref{eq:Mint_Ma}) as
\begin{align}
M^\mu_\ia &=M_c^\mu +T^\mu +
XG_0\big[(M^\mu_u-m^\mu_u)+(M^\mu_t-m^\mu_t)+T^\mu\big]~.
 \label{eq:MiaMc}
\end{align}
In this scheme, therefore, the choice one makes for $M^\mu_c$ (and $T^\mu$)
corresponds to an implicit approximation of the full dynamics contained in the
right-hand side of Eq.~(\ref{eq:Madef}). Moreover, beyond this actual choice,
the only explicit effect of the FSI $X$ is from explicitly \emph{transverse}
loop contributions, which is precisely the same result that was found in
Ref.~\refcite{HH3}. Thus it follows that
\begin{equation}
k_\mu M^\mu_\ia= k_\mu M^\mu_c
\end{equation}
and this approximate interaction current then obviously satisfies the
gauge-invariance condition (\ref{eq:GCMint}).

%%%%%%%%%%%%%%%%%%%%%%%%%%%%%%%%%%%%%%%%%%%%%%%%%%%%%%
\subsection{Phenomenological choice for $M^\mu_c$}

The phenomenological choice that we make here for $M^\mu_c$ is a variant of the
procedure proposed in Refs.~\refcite{HH1,HH2} that is more general than what
was suggested in \refcite{HH3}. We parameterize the $\pi NN$ vertices by
\begin{equation}
F_x = g_\pi \gamma_5 \left[\lambda+(1-\lambda)
\frac{\fs{q}_\pi}{m+m'}\right]f_x~,
 \label{eq:Fdef}
\end{equation}
where $q_\pi$ is the outgoing pion four-momentum, $x=s$, $u$, or $t$ indicate
the kinematic context, $g_\pi$ is the physical coupling constant, $m$ and $m'$
are the nucleon masses before and after the pion is emitted/absorbed and the
parameter $\lambda$ allows for the mixing of pseudoscalar (PS: $\lambda=1$) and
pseudovector (PV: $\lambda=0$) contributions. For simplicity, the functional
dependence $f_x$ of the vertex (which depends on the squared four-momenta of
all three legs) is chosen as common to both PS and PV and it is normalized to
unity if all vertex legs are on-shell. We define then an auxiliary current
\begin{align}
  C^\mu
 &=  -e_\pi\frac{(2q-k)^\mu}{t-q^2}(f_t-\hat{F})
 -e_f \frac{(2p'-k)^\mu}{u-p'^2}(f_u-\hat{F})
 \nonumber\\
 &\qquad\quad\mbox{}
 -e_i \frac{(2p+k)^\mu}{s-p^2}   (f_s-\hat{F})~,
 \label{eq:Cmu}
\end{align}
where
\begin{align}
  \hat{F}= 1-\hat{h}\,\big(1-\delta_s f_s\big) \big(1-\delta_u f_u\big)\big(1-\delta_t
  f_t\big)~.
  \label{eq:Fhatdef}
\end{align}
The factors $\delta_x$ are unity if the corresponding channel contributes to
the reaction in question, and zero otherwise. In principle, the parameter
$\hat{h}$ may be an arbitrary (complex) function, $\hat{h}=\hat{h}(s,u,t)$,
subject to crossing-symmetry constraints.\cite{DW} However, in the application
discussed in the next section, we simply take $\hat{h}$ as a fit constant. With
this choice for $\hat{F}$, the auxiliary current $C^\mu$ is manifestly
 nonsingular, \ie, it is a \emph{contact} current, and in view of charge conservation,
$e_\pi+e_f-e_i=0$, its four-divergence evaluates to
\begin{equation}
  k_\mu C^\mu = e_\pi f_t + e_f f_u - e_i f_s~.
\end{equation}
Using the vertex parametrization (\ref{eq:Fdef}) and writing out the
gauge-invariance condition (\ref{eq:McGI}) explicitly, we may then extract the
ansatz
\begin{align}
  M^\mu_c
  &= -g_\pi \gamma_5 \frac{(1-\lambda)\gamma^\mu}{m'+m}  e_\pi  f_t
 + g_\pi \gamma_5\left[\lambda+  \frac{(1-\lambda)\fs{q}}{m'+m}
 \right]C^\mu
  ~.
  \label{eq:MCbeta}
\end{align}
In this lowest order, therefore, the bare $\gamma_5 \gamma^\mu e_\pi$
Kroll--Ruderman coupling is dressed by the $t$-channel form factor $e_\pi \to
e_\pi f_t$. In addition, there is an auxiliary current given by (\ref{eq:Cmu}),
with the same coupling structure as the underlying $\pi NN$ vertex
(\ref{eq:Fdef}).

Obviously, the above choice of $M^\mu_c$ is not unique, for we can always add
another transverse current to it. We emphasize that such a transverse contact
current in $M^\mu_c$ should not be confused with the transverse contact current
$T^\mu$ appearing in Eq.~(\ref{eq:MiaMc}). In particular, $M^\mu_c$ does not
contribute to the FSI loop integral, but $T^\mu$ does.

%%%%%%%%%%%%%%%%%%%%%%%%%%%%%%%%%%%%%%%%%%%%%%%%%%%%%%
\subsubsection{Next-order approximation}
The approximation made so far was to replace $M^\mu_a$ of (\ref{eq:Madef}) by
(\ref{eq:Ma_approx}), with an undetermined transverse contribution $T^\mu$.
This approximation may be systematically improved by explicitly accounting for
more of the features of $U^\mu$. We mention without derivation that the next
order in this scheme reads
\begin{equation}
  T^\mu = \left(E^\mu -\tilde{E}^\mu\right)G_0 [F\tau] +T'^\mu~,
  \label{eq:Eadd}
\end{equation}
which replaces $T^\mu$ in Eq.~(\ref{eq:Ma_approx}) by an explicit contribution
due to the single-particle exchange current $E^\mu$ (see Fig.~\ref{fig:ucurr}),
still leaving an undetermined transverse current $T'^\mu$. $\tilde{E}^\mu$ is
the $\emph{non-transverse}$ part of $E^\mu$, \ie, one has
\begin{equation}
 k_\mu\left(E^\mu -\tilde{E}^\mu\right) =0
\end{equation}
as an off-shell property, similar to (\ref{eq:Mmdiff}). This scheme can be
systematically extended to all orders of mechanisms contained in $U^\mu$.

%%%%%%%%%%%%%%%%%%%%%%%%%%%%%%%%%%%%%%%%%%%%%%
\begin{figure}[b!]\centering
\includegraphics[width=.85\textwidth,angle=0,clip]{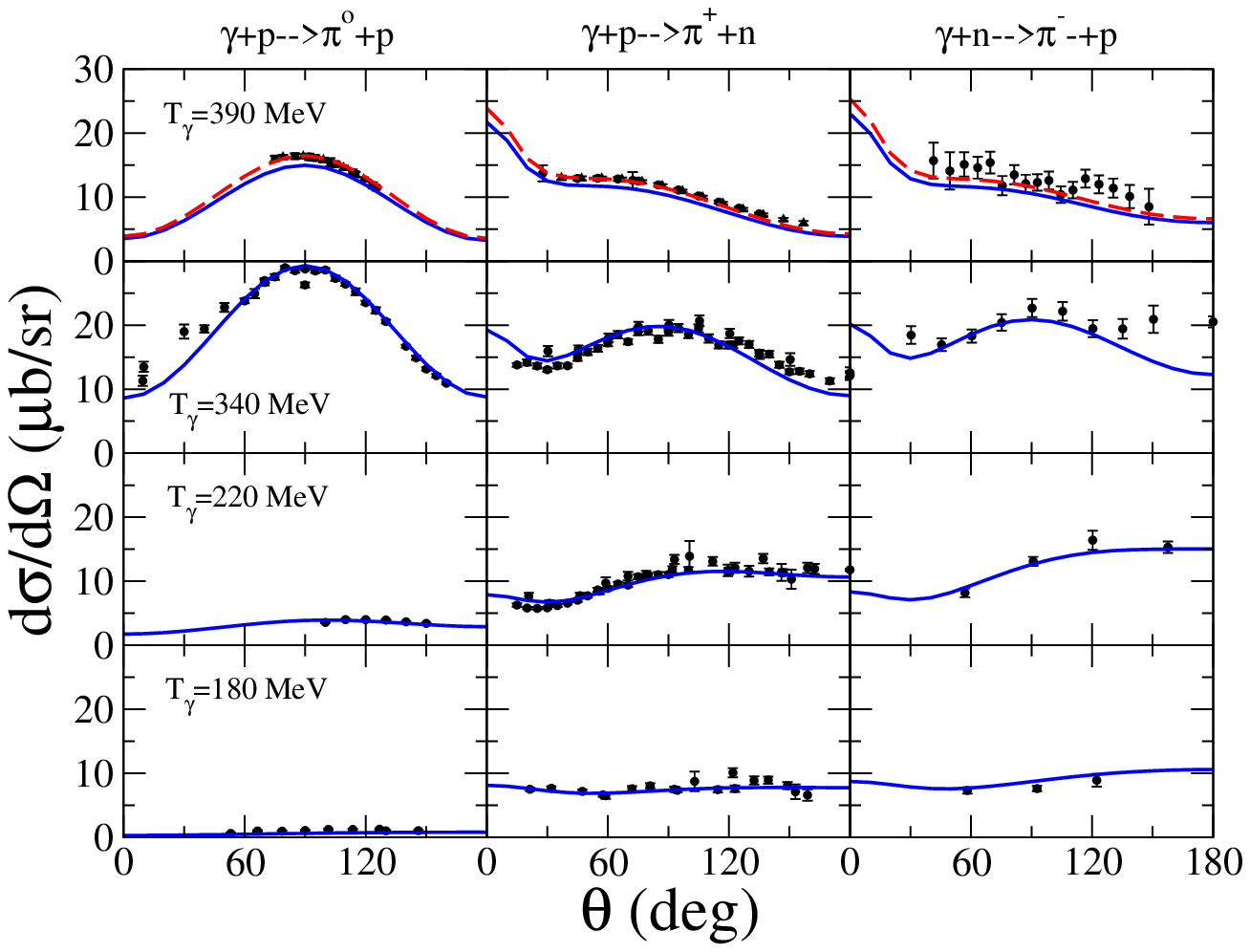}
\parbox{0.5\textwidth}{\mbox{}\\[2ex]
\includegraphics[width=0.55\textwidth,angle=0,clip]{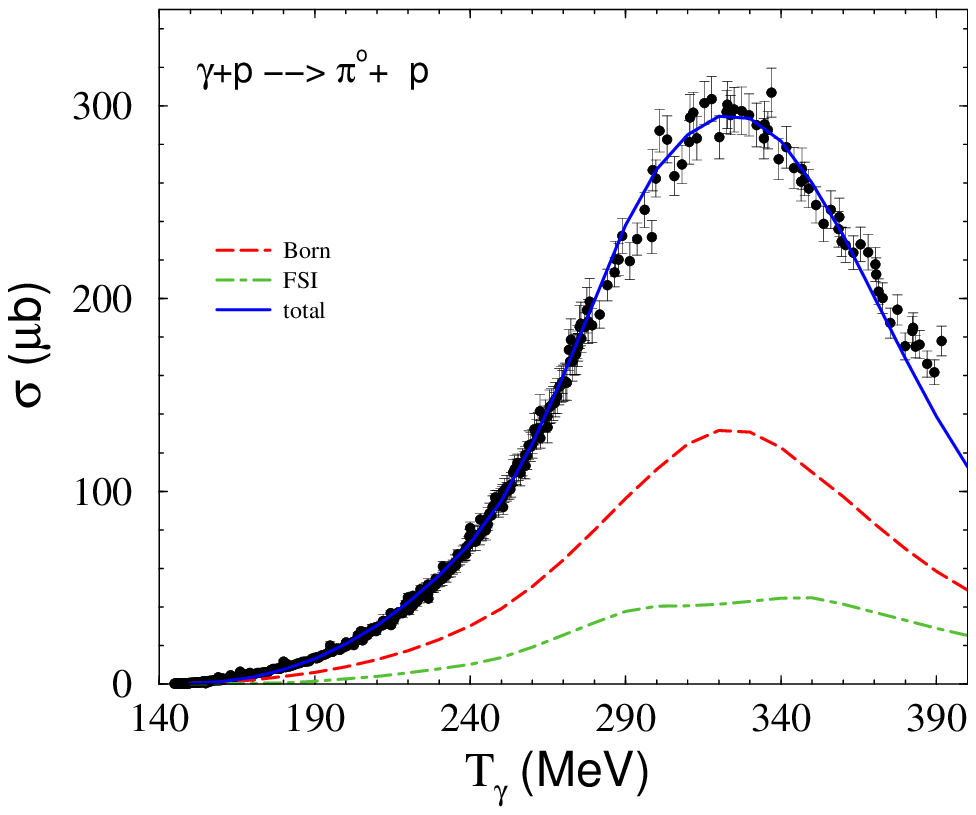}}\hfill
\parbox{.42\textwidth}{%
\caption{\label{fig:datafit}%
Differential and total cross sections. In the top row of the
differential cross sections, the dashed curves correspond to the results
represented by solid curves multiplied by an arbitrary factor of 1.1 (Data:
Refs.~\protect\refcite{Mainz2,dxsc1}.) The total cross section for the
reaction $\gamma + p \rightarrow \pi^0 + p$ is given as a function of photon
incident energy $T_\gamma$. The dashed curve corresponds to the Born
contribution and the dash-dotted curve to the FSI loop contribution. The solid
curve is the total contribution. (Data: Ref.~\protect\refcite{txsc}.)}}
\end{figure}
%%%%%%%%%%%%%%%%%%%%%%%%%%%%%%%%%%%%%%%%%%%%%%

%%%%%%%%%%%%%%%%%%%%%%%%%%%%%%%%%%%%%%%%%%%%%%%%%%%%%%
\subsection{First application: pion photoproduction}

In this section, as a first feasibility study, we apply the approach developed
in the preceding section at its lowest order to the photoproduction reaction
$\gamma + N \to \pi + N$. We restrict ourselves to photon energies up to about
$400$ MeV. Therefore, in addition to the basic nucleons and pions discussed in
the preceding section, our model also incorporates intermediate $\Delta$s in
the $s$- and $u$-channels, and we include the $\rho$, $\omega$, and $a_1$
meson-exchanges in the $t$-channel. Note here that transition currents between
different hadronic states are transverse individually and therefore play no
role for the issue of gauge invariance. For the $\pi N$ FSI, we employ the $\pi
N$ $T$-matrix developed by the J\"ulich group.\cite{Oliver} Full details of
this first application will be given elsewhere. We only mention here that the
present model has only three free parameters that are adjusted to reproduce the
pion photoproduction cross section data, as shown in Fig.~\ref{fig:datafit}.

Generally, for the total cross section, the agreement with the data is very
good except for energies above $T_\gamma \sim 360$ MeV, where the prediction
tends to underestimate the data. In particular, around $T_\gamma \approx 390$
MeV, the discrepancy is about $10\%$. We also see that the FSI loop
contribution is relatively small compared to the Born contribution. However, it
plays a crucial role in reproducing the observed energy dependence through its
interference with the dominant Born term. For differential cross sections for
neutral and charged pion productions at various energies, we also see that,
overall, the data are reproduced quite well. The dashed curves in the top row
correspond to the results represented by the solid curves multiplied by an
arbitrary factor of 1.1 to facilitate visualizing that the shape of the angular
distribution is well reproduced in spite of the absolute normalization being
underestimated at this energy.

%%%%%%%%%%%%%%%%%%%%%%%%%%%%%%%%%%%%%%%%%%%%%%%%%%%%%%
\section{Summary}

By exploiting the generalized Ward-Takahashi identity for the production
amplitude and total charge conservation, we have constructed a fully gauge
invariant (pseudoscalar) meson photoproduction amplitude which includes the
hadronic final-state interaction explicitly. The method based on an earlier
field-theoretical approach\cite{HH1} is quite general and can be readily
extended to any other meson photo- and electroproduction reactions. This method
should be particularly relevant for the latter reaction. As an example of
application of the present approach in its lowest order, we have calculated
both the neutral and charged pion photoproduction processes off nucleons up to
about $400$ MeV photon incident energy which illustrates the feasibility of the
present method. Obviously, for a more quantitative calculation, including not
only cross sections but also other observables, some of the approximations made
in the present feasibility study need to be improved.

%%%%%%%%%%%%%%%%%%%%%%%%%%%%%%%%%%%%%%%%%%%%%%%%%%%%%%
\section*{Acknowledgments}
This work is partly supported by the Forschungszentrum J\"ulich, COSY Grant
No.\ 41445282\;(COSY-58).

\end{document}